\documentclass[
aps,%
11pt,%
final,%
notitlepage,%
oneside,%
onecolumn,%
nobibnotes,%
nofootinbib,%
superscriptaddress,%
noshowpacs,%
centertags]%
{revtex4}

\usepackage{multirow}
\renewcommand{\baselinestretch}{1.0}

\begin{document}

\selectlanguage{english}

%\inEnglish{%
%\OrigPages{399--412} \OrigCopyrightedAuthors{F.G.~Kopylova,
%A.I.~Kopylov}}

%\toctitle {Peculiar Motions of Galaxy Clusters in the Regions of the

\title{Peculiar Motions of Galaxy Clusters in the Regions of Superclusters of Galaxies}

\author{\firstname{F.~G.}~\surname{Kopylova}}
%\email{flera@sao.ru}
%\firstaffiliation{\saoname}

\author{\firstname{A.~I.}~\surname{Kopylov}}
\affiliation{\saoname}

%\received{May 11, 2017} \revised{August 9,2017}
\onecolumngrid
\begin{center}
{\scriptsize
Original Russian Text @ F.G.~Kopylova, A.I.~Kopylov,
published in Astrofizicheskii Byulleten, 2017,\\
Vol.72, No.4, pp.399-412}
\end{center}

\begin{abstract}
We present results of the study of peculiar motions of 57 clusters
and groups of galaxies in the regions of the Corona Borealis (CrB),
Bootes (Boo), Z\,5029/A\,1424, A\,1190, A\,1750/A\,1809 superclusters
of galaxies and 20 galaxy clusters located beyond massive structures
($0.05<z<0.10$). Using the SDSS (Data Release 8) data, a sample of
early-type galaxies was compiled in the systems under study, their
fundamental planes were built, and relative distances and peculiar
velocities were determined. Within the galaxy superclusters,
significant peculiar motions along the line of sight are observed
with rms deviations of $652\pm50$~km\,s$^{-1}$---in CrB, \mbox
{$757\pm70$}~km\,s$^{-1}$---in Boo. For the most massive A\,2065
cluster in the CrB supercluster, no peculiar velocity was found.
Peculiar motions of other galaxy clusters can be caused by their
gravitational interaction both with A\,2065 and with the A\,2142
supercluster. It has been found that there are two superclusters
projected onto each other in the region of the Bootes supercluster
with a radial velocity difference of about 4000~km\,s$^{-1}$. In the
Z\,5029/A\,1424 supercluster near the rich Z\,5029 cluster, the most
considerable peculiar motions with a rms deviation of
$1366\pm170$~km\,s$^{-1}$ are observed. The rms deviation of
peculiar velocities of 20 clusters that do not belong to large-scale
structures is equal to  $0\pm20$~km\,s$^{-1}$. The whole sample of
the clusters under study has the mean peculiar velocity equal to
$83\pm130$~km\,s$^{-1}$ relative to the cosmic microwave background.
\end{abstract}


\footnotetext{email: flera@sao.ru}
\maketitle

\section{INTRODUCTION}
In the large-scale structure of the Universe, the largest systems of
galaxies, superclusters, are connected with each other with filaments
from galaxies and are located at the boundaries of voids. Such large
systems (or their separate parts) can be gravitationally bound if the
contrast of the mean density of galaxies within is
sufficient (see, e.g.,~\cite{Dunner:Kopylova_n_en}). If galaxy clusters in
superclusters are quite densely located in the projection to the
celestial sphere, one can expect the presence of large peculiar
velocities in them which can be measured by determining distances to
galaxy clusters. Such compact superclusters within $0.05<z<0.1$ are
superclusters of Corona Borealis (CrB), Ursa Major (UMa) and Bootes
(Boo). The most well-studied are the CrB and UMa superclusters and
the Boo supercluster is the least studied.

As a result of the detailed study of seven galaxy clusters  A\,2061,
A\,2065, A\,2067, A\,2079, A\,2089, A\,2092, A\,2124 in
CrB~\cite{Postman:Kopylova_n_en}, it was found that with the radial velocity
dispersion equal to  1300~km\,s$^{-1}$ and the size equal to
$18.6$~Mpc, the supercluster mass is $1.2\times 10^{16}~M_\odot$.
This mass is sufficient for the system to be gravitationally bound
and peculiar velocities not to exceed 2200~km\,s$^{-1}$.
Small~et~al.~\cite{Small1:Kopylova_n_en}, having considerably increased the number
of galaxies with the measured redshifts and based on modeling a
supercluster with parameters  similar to those of CrB, determined the
mass of the supercluster equal to $4.3\times 10^{16}~M_\odot$ and
proved the hypothesis presented by Postman in paper~\cite{Postman:Kopylova_n_en} on
the fact that the supercluster is gravitationally bound. They also
supposed that the supercluster is in the dynamic phase of initial
gravitational collapse. In paper~\cite{Kopylova1:Kopylova_n_en}, we tried to
show the validity of the latest assumption having measured the
distances of the galaxy clusters belonging to the supercluster using
the Kormendy relation~\cite{Kormendy:Kopylova_n_en}. Recent studies of peculiar
motions of galaxy clusters with the fundamental plane (FP) and the
Kormendy relation (SDSS and DR7 data) revealed in CrB great peculiar
velocities indicating the galaxy clusters to be gravitationally
bound~\cite{Batiste:Kopylova_n_en}. In~\cite{Pearson:Kopylova_n_en}, the mass of the supercluster
has been estimated with the caustics method, the virial theorem, and
model calculations (the spherical collapse model); it has been also
determined that the central region of the supercluster (A\,2056,
A\,2061, A\,2065, A\,2067, and A\,2089) with a mass of about $1.43
\times 10^{16}~M_{\odot}$ is gravitationally bound, has reached the
turnaround radius, and is in the state of collapse.

A study of the  gravitational potential distribution from the SDSS
(DR10) data in the CrB region revealed two deep potential wells that
correspond to galaxy clusters systems at \mbox {$z\sim 0.074$} and
\mbox {$z\sim 0.09$}: one system is near the CrB supercluster, in
another system there is the very rich A\,2142
cluster~\cite{Pillastrini:Kopylova_n_en}. The mass of the system with A\,2142
determined with the virial theorem 1.4~times exceeds the mass of the
system with CrB. It was found that these galaxy cluster systems are
connected with a filament  from galaxies and are probably
gravitationally bound. From the SDSS (DR8, DR10) data, in the region
of the A\,2142 cluster a supercluster has been found consisting of 14
galaxy systems (with at least ten members) which has the estimated
mass $M_{\rm est}=6.2\times 10^{15}~M_{\odot}$~\cite{Einasto2:Kopylova_n_en}. The
supercluster found apparently is a part of a larger system at $z\sim
0.09$.

A similar method applied to another massive supercluster, Shapley,
allowed one to find that the central core of Shapley consisting of
two galaxy clusters is also in the state of the initial gravitational
collapse~\cite{Reisenegger:Kopylova_n_en}.

Model calculations of peculiar motions of galaxy clusters (taking
errors into account) for the CDM model with $\Omega=0.3$ have showed
that the distribution of one-dimensional peculiar velocities of the
clusters has a peak at 400~km\,s$^{-1}$ and stretches to
500~km\,s$^{-1}$. Their rms deviations is $ \langle
V^2_{1D}\rangle^{1/2}=544$~km\,s$^{-1}$ and in observations it is
$725\pm50$~km\,s$^{-1}$~\cite{Bahcall:Kopylova_n_en}. It is also noted in the paper
that despite the obtained peculiar velocities higher than
2000~km\,s$^{-1}$, the observed data are in agreement with model
predictions at a level of about~$1$--$3\sigma$.

The Tully-Fisher relation between the parameters of spiral galaxies
allowed the authors of paper~\cite{Masters:Kopylova_n_en} to find the rms
deviation of peculiar velocities along the line of sight for 31
galaxy clusters to be equal to $368\pm40$~km\,s$^{-1}$ (for
``In''-sample of the galaxy cluster members) which is close to model
calculations.

Using results of the study of the structure, dynamics, and peculiar
velocities of galaxy clusters (\mbox{$z=0.22$}, SDSS
DR10~\cite{O'Mill:Kopylova_n_en}) belonging to the\linebreak SC\,0028\mbox{-}0005
supercluster, we found the estimate of the rms deviation of
one-dimensional peculiar velocities with squared consideration of
errors to be equal to approximately $540\pm50$~km\,s$^{-1}$ (as it
was not done in the paper indicated). Peculiar velocities there were
determined with the FP of early-type galaxies.

Determination of the spatial structure and peculiar motions in the
region of the Ursa Major supercluster  (SDSS DR4) showed that the
supercluster has the dimensions \mbox {$40\times15$}~Mpc in the
plane of sky and looks compact~\cite{Kopylova2:Kopylova_n_en}. It consists of
three filamentary structures which do not intersect and influence each
other; thus, no considerable peculiar motions are observed in the
supercluster: the rms deviation of the observed peculiar velocities
is \mbox {$\langle V^2 \rangle^{1/2} = 290\pm120$}~km\,s$^{-1}$
taking into consideration the measurement errors~\cite{Kopylova2:Kopylova_n_en}.
Peculiar motions are also determined with the FP of early-type
galaxies.

We have studied peculiar motions of 26 groups and clusters of the
Hercules and Leo superclusters of galaxies\linebreak \mbox {($0.027<z<0.045$)} using
the SDSS (DR8) data and the FP of early-type galaxies. We have
found that for the superclusters under study  the Hubble's law is
satisfied, and within them considerable peculiar velocities of the
galaxy clusters along the line of sight are observed, with the rms
deviations \mbox{$736\pm50$~km\,s$^{-1}$} (Hercules) and
$625\pm70$~km\,s$^{-1}$ (Leo)~\cite{Kopylova3:Kopylova_n_en}. These
results were obtained with the forward FP without taking into
consideration the Malmquist bias.

The aim of the present paper is the study of peculiar motions of
groups and clusters of galaxies located in the regions of the massive
superclusters of the  Corona Borealis and  A\,2142, Bootes,
in the regions of smaller superclusters of the galaxies such as
SCl\,24, SCl\,38, SCl\,61~\cite{Einasto1:Kopylova_n_en}, and pairs of galaxy
clusters. Moreover, there are 20 clusters of galaxies in our sample
that do no not practically belong to such large superclusters at the
redshifts $0.05<z<0.10$. These clusters can form pairs or small
superclusters but at least they have no great satellites within $cz$
$\pm3000$~km\,s$^{-1}$ and at a distance of 4~Mpc in the plane of sky.
Along with this, the A\,1035A and A\,1035B clusters are not
gravitationally bound systems~\cite{Kopylov1:Kopylova_n_en}. In the A\,1691,
A\,1024, and Anon1 clusters, there are small groups of galaxies
falling onto them. We conducted the work using the data from the SDSS
(Sloan Digital Sky Survey, DR7, DR8), NED catalogs.

The paper is arranged in the following way. In Section~2, we describe
the selection of early-type galaxies, build the FP, and determine
relative distances of groups and clusters of galaxies. In Section~3,
we calculate peculiar velocities and determine their rms deviations
in the superclusters. In the Conclusion, we give the results
obtained. In this paper, we used the following cosmological
parameters: \mbox{$\Omega_m=0.3$,} \mbox{$\Omega_{\Lambda}=0.7$,}
\mbox{$H_0=70$~km\,s$^{-1}$\,Mpc$^{-1}$}.

\section{BUILDING THE FUNDAMENTAL PLANE OF EARLY-TYPE GALAXIES}
\subsection{Description of the Sample}

%1
\renewcommand
\baselinestretch{0.85}
\begin{table*}[]
\setcaptionmargin{0mm} \onelinecaptionstrue \captionstyle{normal}
\caption{Position of galaxy systems in the sky and their dynamic
properties} \label{data1:Kopylova_n_en}
\medskip
\begin{tabular}{l|c|c|c|c|c|r@{$\,\pm\,$}l|r@{$\,\pm\,$}l}
\hline
Cluster & $\alpha$ (J2000), & $\delta$(J2000), & $z_{\rm spec}$ & $R_{200}$, & $N_z$ & \multicolumn{2}{c|}{$\sigma$,}     & \multicolumn{2}{c}{$M_{200}$,}\\
%\hline
                                              & hh\,mm\,ss.s      & dd\,mm\,ss       &                                 & Mpc        &                         & \multicolumn{2}{c|}{km\,s$^{-1}$} & \multicolumn{2}{c}{$10^{14}~M_{\odot}$} \\
\hline
SCl CrB             &            &             &        &      &     & \multicolumn{2}{c|}{} & \multicolumn{2}{c}{} \\
  A\,2019           & 15 03 28.9 & +27 09 16   & 0.0817 & 0.82 & 14  & $345$&$92$  &  $ 0.68$&$0.50$ \\
  A\,2056           & 15 19 02.8 & +28 20 32   & 0.0754 & 0.57 & 5   & $237$&$106$ &  $ 0.22$&$0.30$ \\
  A\,2061           & 15 21 20.6 & +30 40 15   & 0.0782 & 1.70 & 121 & $712$&$65$  &  $ 6.03$&$1.65$ \\
  A\,2065           & 15 22 29.2 & +27 42 27   & 0.0726 & 2.64 & 210 & $1104$&$76$ &  $22.53$&$4.29$ \\
  A\,2067           & 15 23 02.0 & +30 52 39   & 0.0733 & 0.68 & 14  & $286$&$76$  &  $ 0.39$&$0.31$ \\
  A\,2079           & 15 23 45.1 & +28 55 43   & 0.0661 & 1.48 & 85  & $618$&$67$  &  $ 3.96$&$1.29$ \\
  A\,2089           & 15 32 49.8 & +28 02 22   & 0.0739 & 1.27 & 51  & $531$&$74$  &  $ 2.50$&$1.05$ \\
  A\,2092           & 15 33 15.4 & +31 08 42   & 0.0669 & 1.17 & 37  & $486$&$80$  &  $ 1.93$&$0.88$ \\
  A\,2124           & 15 44 59.0 & +36 06 34   & 0.0660 & 1.77 & 85  & $736$&$80$  &  $ 6.70$&$2.18$ \\
CL\,1529+29$^*$     & 15 30 28.4 & +28 57 03   & 0.0845 & 0.98 & 19  & $411$&$94$  &  $ 1.16$&$0.80$ \\
\hline
A\,2142             & 15 58 20.0 & +27 14 00   & 0.0904 & 2.28 & 190 & $963$&$70$   &  $14.82$&$3.23$ \\
\hline
SCl Bootes          &            &             &        &      &     & \multicolumn{2}{c|}{} & \multicolumn{2}{c}{} \\
A\,1775A            & 13 42 42.0 & +26 14 23   & 0.0656 & 0.78 & 18  & $324$&$76$  &  $ 0.57$&$0.40$ \\
A\,1781$^*$         & 13 44 45.5 & +29 44 44   & 0.0630 & 0.87 & 25  & $362$&$72$  &  $ 0.80$&$0.48$ \\
A\,1795             & 13 48 52.5 & +26 35 34   & 0.0632 & 1.86 & 125 & $775$&$70$  &  $ 7.83$&$2.12$ \\
A\,1825             & 13 58 03.3 & +20 37 08   & 0.0641 & 1.52 & 40  & $633$&$100$ &  $ 4.26$&$2.02$ \\
A\,1828             & 13 58 14.7 & +18 20 46   & 0.0634 & 0.80 & 13  & $335$&$93$  &  $ 0.63$&$0.52$ \\
A\,1831A            & 13 59 10.7 & +27 56 26   & 0.0637 & 1.15 & 34  & $480$&$82$  &  $ 1.86$&$0.95$ \\
A\,1775B            & 13 41 49.1 & +26 22 24   & 0.0759 & 1.39 & 62  & $581$&$74$  &  $ 3.28$&$1.25$ \\
A\,1800             & 13 49 23.6 & +28 06 26   & 0.0761 & 1.68 & 67  & $705$&$86$  &  $ 5.86$&$2.14$ \\
A\,1831B            & 13 59 15.1 & +27 58 34   & 0.0762 & 2.27 & 89  & $952$&$101$ &  $14.43$&$4.59$ \\
A\,1898$^*$         & 14 20 38.9 & +25 15 28   & 0.0792 & 1.04 & 21  & $434$&$95$  &  $ 1.36$&$0.89$ \\
CL\,1350+29         & 13 50 15.5 & +29 13 17   & 0.0772 & 0.86 & 23  & $359$&$75$  &  $ 0.77$&$0.48$ \\
\hline
SCl Z\,5029/A\,1424 &            &             &        &      &     & \multicolumn{2}{c|}{} & \multicolumn{2}{c}{} \\
A\,1424             & 11 57 29.0 & +05 0 5 21  & 0.0771 & 1.51 & 63  & $632$&$80$  &  $ 4.22$&$1.60$ \\
A\,1516             & 12 18 52.4 & +05 14 44   & 0.0773 & 1.58 & 61  & $660$&$84$  &  $ 4.80$&$1.83$ \\
Z\,4905             & 12 10 16.8 & +05 23 10   & 0.0780 & 1.36 & 42  & $568$&$88$  &  $ 3.06$&$1.42$ \\
Z\,5029             & 12 17 41.1 & +03 39 21   & 0.0785 & 2.18 & 126 & $912$&$81$  &  $12.67$&$3.38$ \\
\hline
SCl A\,1190         &            &             &        &      &     & \multicolumn{2}{c|}{} & \multicolumn{2}{c}{} \\
    A\,1173         & 11 09 15.3 & +41 33 41   & 0.0770 & 1.23 & 35  & $516$&$87$  &  $ 2.30$&$1.16$ \\
    A\,1190         & 11 11 43.6 & +40 49 15   & 0.0762 & 1.60 & 79  & $670$&$75$  &  $ 5.03$&$1.69$ \\
    A\,1203         & 11 13 48.2 & +40 17 09   & 0.0761 & 0.99 & 42  & $416$&$64$  &  $ 1.20$&$0.55$ \\
\hline
SCl A\,1750/A\,1809 &            &             &        &      &     & \multicolumn{2}{c|}{} & \multicolumn{2}{c}{} \\
A\,1750             & 13 30 50.6 & $-01~51~43$ & 0.0869 & 1.78 & 93  & $747$&$77$  &  $ 6.93$&$2.14$ \\
A\,1773             & 13 42 09.6 &  +02 13 38  & 0.0784 & 1.98 & 83  & $832$&$91$  &  $ 9.62$&$3.16$ \\
A\,1780             & 13 44 40.6 &  +02 51 43  & 0.0788 & 1.13 & 37  & $474$&$78$  &  $ 1.78$&$0.88$ \\
A\,1809             & 13 53 06.4 &  +05 08 59  & 0.0802 & 1.74 & 84  & $729$&$80$  &  $ 6.46$&$2.13$ \\
\hline
A\,0602             & 07 53 26.6 &  +29 21 34  & 0.0612 & 1.35 & 59  & $560$&$67$  &  $ 2.96$&$1.16$ \\
A\,0671             & 08 28 31.6 &  +30 25 52  & 0.0505 & 1.95 &116  & $805$&$75$  &  $ 8.83$&$2.47$ \\
A\,1024             & 10 28 23.5 &  +03 45 32  & 0.0745 & 1.38 & 42  & $578$&$89$  &  $ 3.23$&$1.49$ \\
$A\,1035A^*$        & 10 32 19.4 &  +40 10 10  & 0.0688 & 1.35 & 52  & $563$&$78$  &  $ 2.99$&$1.24$ \\
A\,1035B            & 10 32 14.0 &  +40 16 16  & 0.0790 & 1.46 & 37  & $613$&$101$ &  $ 3.85$&$1.90$ \\
$A\,1066^*$         & 10 39 25.1 &  +05 10 15  & 0.0701 & 1.84 & 95  & $768$&$79$  &  $ 7.60$&$2.50$ \\
A\,1205             & 11 13 21.4 &  +02 32 39  & 0.0773 & 1.88 & 74  & $787$&$91$  &  $ 8.15$&$2.83$ \\
A\,1238             & 11 22 54.3 &  +01 06 52  & 0.0753 & 1.29 & 61  & $541$&$69$  &  $ 2.65$&$1.01$ \\
A\,1371             & 11 45 20.6 &  +15 29 28  & 0.0700 & 1.32 & 45  & $552$&$82$  &  $ 2.82$&$1.26$ \\
A\,1589             & 12 41 17.5 &  +18 34 28  & 0.0725 & 1.86 &109  & $778$&$74$  &  $ 7.89$&$2.25$ \\
A\,1668             & 13 03 46.6 &  +19 16 17  & 0.0647 & 1.52 & 63  & $635$&$80$  &  $ 4.31$&$1.63$ \\
A\,1691             & 13 11 08.6 &  +39 13 37  & 0.0733 & 1.76 & 82  & $743$&$82$  &  $ 6.87$&$2.83$ \\
A\,1767             & 13 36 08.3 &  +59 12 23  & 0.0707 & 1.95 &121  & $816$&$74$  &  $ 9.11$&$2.48$ \\
A\,1904             & 14 22 10.2 &  +48 34 15  & 0.0721 & 1.84 &113  & $771$&$72$  &  $ 7.68$&$2.18$ \\
A\,1991             & 14 54 31.5 &  +18 38 32  & 0.0591 & 1.33 & 79  & $554$&$62$  &  $ 2.87$&$0.96$ \\
A\,2029             & 15 10 56.1 &  +05 44 41  & 0.0785 & 2.50 &180  & $1046$&$78$ &  $23.58$&$4.28$ \\
A\,2033             & 15 11 26.5 &  +06 20 58  & 0.0806 & 2.03 & 67  & $853$&$104$ &  $10.35$&$3.78$ \\
A\,2064             & 15 20 52.2 &  +48 39 38  & 0.0740 & 1.51 & 40  & $633$&$100$ &  $ 4.24$&$2.01$ \\
A\,2244             & 17 02 42.5 &  +34 03 36  & 0.0989 & 2.48 & 95  & $1049$&$108$&  $19.08$&$5.89$ \\
A\,2245             & 17 02 33.1 &  +33 31 00  & 0.0879 & 2.46 &112  & $1037$&$98$ &  $18.53$&$5.25$ \\
RXCJ1022            & 10 22 10.3 &  +38 31 04  & 0.0550 & 1.33 & 59  & $551$&$72$  &  $ 2.83$&$1.11$ \\
RXCJ1351            & 13 51 41.9 &  +46 22 00  & 0.0634 & 1.24 & 51  & $517$&$72$  &  $ 2.32$&$0.97$ \\
SHK352              & 11 21 32.6 &  +02 53 14  & 0.0504 & 1.29 & 63  & $532$&$67$  &  $ 2.55$&$0.96$ \\
Z6718               & 14 21 35.8 &  +49 33 04  & 0.0718 & 1.32 & 27  & $550$&$106$ &  $ 2.79$&$1.61$ \\
\hline
\end{tabular}
\end{table*}
\renewcommand
\baselinestretch{1.0}

Our sample is compiled from 57 galaxies having the redshifts $z >
0.05$. A number of galaxy clusters is located in the regions of the CrB
and Bootes superclusters, several systems are in the regions of the
superclusters that are less rich: Z\,5029/A\,1424 (SCl\,24), A\,1190
(SCl\,38), A\,1750/A\,1809 (SCl\,61) (the superclusters are named by
the richest cluster in them, their names given in brackets are
from~\cite{Einasto3:Kopylova_n_en}). We included two pairs of galaxy systems
A\,2244+A\,2245  and A\,2029+A\,2033 into the sample; the rest 20
galaxy clusters do not belong to large-scale structures but form the
field of galaxy systems. Estimates of dynamic characteristics of
galaxy systems are based on measurements of one-dimensional
dispersion of radial velocities from which the virial mass is
calculated within the empirical radius $R_{200}$ on the
assumption of $M(r)\propto r$. The radius $R_{200}$ is close
(although smaller) to the virial and within its the density of
the system 200~times exceeds the critical density of the Universe.
$R_{200}$ can be estimated with the formula \mbox {$R_{200} = \sqrt
{3} \sigma /10H(z)$}~Mpc~\cite{Carlberg:Kopylova_n_en}.

On the assumption of $M_{200} \simeq M_{\rm vir}$, the mass within
$R_{200}$ is $M_{200} = 3G^{-1}R_{200}\sigma_{200}^{2}$. First, we
estimated the average radial velocity of the cluster $cz$  and its
dispersion $\sigma$, then from the dispersion---the radius $R_{200}$.
Galaxies with the velocities greater $2.7\sigma$ were considered as
background. For the other galaxies, we determined $cz$, $\sigma$, and
$R_{200}$ again. In an iterative way, we found all the
characteristics of the galaxy clusters within the radius given. As a
rule, we took the brightest galaxy as a center of galaxy cluster,
near which the center of X-ray emission was situated. Galaxy clusters
that do not belong to large-scale superclusters like CrB and Boo have
\mbox{$\sigma > 500$~km\,s$^{-1}$}, and all the galaxy clusters
belonging to in CrB and Boo are taken including those with smaller
dispersion. Table~\ref{data1:Kopylova_n_en} presents the measured dynamic
parameters of galaxy clusters for the radius $R_{200}$: name of the
cluster, center coordinates, heliocentric redshifts in the CMB system (the
corrections taken from the NED database), the radius  $R_{200}$ in
Mps, the number of galaxies with the measured radial velocities
\mbox{$r_{\rm Pet}<17\fm77$}), radial velocity dispersion with
cosmological correction  $(1+z)^{-1}$ and error, the mass $M_{200}$
with error corresponding to determination error $\sigma$. For galaxy
clusters marked with an asterisk, we took a centroid as a center,
because several bright galaxies are situated in their central parts.

\subsection{Comments to the Sample}

{\it The Corona Borealis supercluster}. The first study of the
dynamics of galaxy clusters in the region of the CrB supercluster was
based on six systems:  A\,2061, A\,2067, A\,2065, A\,2079, A\,2089,
and A\,2092~\cite{Postman:Kopylova_n_en}. For a more detailed study
in this field, Small with his colleagues in the Norris
Survey~\cite{Small1:Kopylova_n_en, Small2:Kopylova_n_en,
Small3:Kopylova_n_en} included two more systems in the supercluster:
A\,2056 and CL\,1529+29. In the galaxy superclusters
catalog~\cite{Einasto3:Kopylova_n_en}, A\,2019 and A\,2124 were added
to galaxy clusters~\cite{Postman:Kopylova_n_en}. In
paper~\cite{Kopylova1:Kopylova_n_en}, the authors used eight galaxy
clusters from catalog~\cite{Postman:Kopylova_n_en} and the Kormendy
relation for early-type galaxies in order to determine the dynamic
state of the CrB supercluster. To determine the dynamic state of the
supercluster with the FP of early-type galaxies using the data from
the SDSS (DR7) in~\cite{Batiste:Kopylova_n_en}, six galaxy systems
were  studied in paper~\cite{Batiste:Kopylova_n_en} as well as in
paper~\cite{Postman:Kopylova_n_en}, and then
in~\cite{Pearson:Kopylova_n_en}---eight, as well as
in~\cite{Small1:Kopylova_n_en}.

\begin{figure*}[]
\setcaptionmargin{5mm} \onelinecaptionsfalse
\includegraphics[scale=0.5,angle=0]{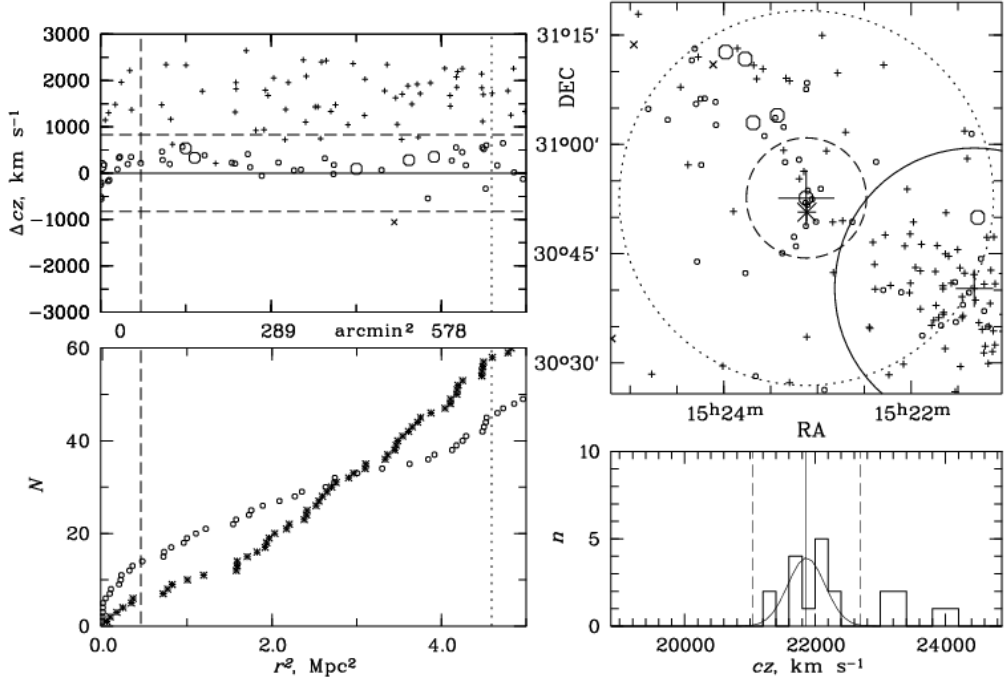}
\captionstyle{normal} \caption{Distribution of the galaxies in the
cluster A\,2067 ($r_{\rm Pet}<17\fm77$). The left-hand panel shows
the deviation of the radial velocities of the galaxies  from the mean
radial velocity of the cluster determined from galaxies within the
radius $R_{200}$. Horizontal dashed lines correspond to deviations of
$\pm2.7\sigma$, a vertical dotted line---the Abell radius $R_{A}$
(2.14~Mpc). Larger-sized circles denote galaxies brighter than $M_K <
-24\fm5$, pluses---background galaxies (from the A\,2061 cluster),
crosses---foreground galaxies. The lower left-hand panel gives the
integrated distribution of the full number of galaxies depending on
the squared distance from the cluster center. Circles correspond to
galaxies marked with circles in the left-hand figure, asterisks---to
the background galaxies. A dashed and dotted lines denote the same as
in the upper panel. The upper right-hand panel shows the distribution
of galaxies given in the upper left-hand figure in the sky in the
equatorial coordinate system. Circles show the regions with the
radius $R_{200}$, $R_A$. The legend is the same. The field of study
is limited with $25\farcm5$. A big cross indicates the center of the
A\,2067 cluster, an asterisk below marks the center of the X ray
emission in accordance with~\cite{Ebeling:Kopylova_n_en}. In the right-hand corner,
a solid line shows the region with the radius  $R_{200}$ of the
A\,2061 cluster, the cross marks its center. The lower right-hand
panel presents the distribution of all the galaxies within the radius
$R_{200}$ by radial velocities. A Gaussian corresponding to $\sigma$
of the cluster is shown with a solid line for the cluster members. A
solid vertical line indicates the mean radial velocity of the
cluster, dashed lines correspond to the deviations $\pm2.7\sigma$. }
\label{clus1:Kopylova_n_en}
\end{figure*}

In order to obtain new measurements of peculiar velocities, at first,
we determined dynamic characteristics of galaxy clusters
themselves with the data from the SDSS (DR7), because in the region
of A\,2067, for example, we do not find such a rich galaxy system in
contrast with many other authors. For a more detailed
characterization of the structure and kinematics of the A\,2067
cluster and its vicinity, Fig.~\ref{clus1:Kopylova_n_en} shows the following as an
example:
%\begin{minipage}{1.0\linewidth}
\begin{list}{}{
\setlength\leftmargin{5mm} \setlength\topsep{2mm}
\setlength\parsep{0mm} \setlength\itemsep{2mm} }
%\item 1)
\item [a)] deviation of the radial velocities of the
galaxies--members of the cluster and galaxies referring to the
background from the mean radial velocity of the cluster depending on
the squared radius (distance from the cluster center);
%\item 2)
\item [b)] integrated distribution of the number of galaxies
depending on the squared radius;
%\item 3)
\item [c)] position of galaxies in the sky plane in equatorial
coordinates;
%\item 4)
\item [d)] histogram of radial velocity distributions of all the
galaxies with the radius $R_{200}$.
\end{list}
%\end{minipage}

%Fig.1

In the upper right-hand panel of Fig.~\ref{clus1:Kopylova_n_en}, the big crosses
correspond to the centers of the clusters A\,2061 (in the lower
corner) and A\,2067 (in the center) and the radii $R_{200}$ that we
found are shown with the solid and dashed liens respectively. It can
be noticed that  A\,2067 is a small group in the filament stretching
from the rich A\,2061 cluster. As far as no X-ray emission was
detected in the system, it is not a random projection of galaxies but
began to separate  from the filament into a single group (the upper
left-hand panel of Fig.~\ref{clus1:Kopylova_n_en}). The A\,2056 system of galaxies
is also a small group, although, without any detected emission. The
CL\,1529+29 galaxy system is a binary non-virialized system with two
brightest galaxies of approximately similar brightness. The centroid
of galaxies was taken as a center of this system.

{\it The Bootes supercluster}. In the galaxy supercluster
catalog~\cite{Einasto3:Kopylova_n_en}, the Bootes supercluster is specified as
consisting of 12 members. The SDSS data show that two clusters from
the Abell catalog:  A\,1861 and A\,1927 are more remote systems, and
two other A\,1827 and A\,1873 are very poor groups which we did not
considered in our study. Furthermore, we have derived that the
A\,1775 and A\,1831 clusters are bimodal galaxy
clusters~\cite{Kopylov2:Kopylova_n_en,Kopylov3:Kopylova_n_en}, where main clusters which we have
denoted as A\,1775B and A\,1831B (as they are more distant) have the
detected X-ray emission. The region of the the Bootes supercluster is
a layered structure: in the nearest layer with $z\sim 0.064$ we have
found six galaxy systems, among which the A\,1795 cluster is the
richest; in the distant layer with $z\sim 0.077$---five galaxy
systems with the richest one---A\,1831B. The A\,1825 and A\,1828
clusters in the nearby system  are situated slightly aside from the
main sample and the A\,1898 cluster---aside from the distant sample.

{\it The Z\,5029/A\,1424 supercluster} is a part  of the SCl\,24
supercluster. In the catalog of galaxy superclusters~\cite{Einasto1:Kopylova_n_en},
two Abell clusters, A\,1424 and A\,1516, are referred to this system,
although, the most massive system in this region is the Z\,5029
cluster (in Fig.~10 in paper~\cite{Einasto1:Kopylova_n_en} this cluster is marked
as A\,1516 by mistake). The A\,1516 cluster is located in the
sky plane above Z\,5029 and there is the Z\,4905 cluster near it.
According to the obtained dynamic parameters of these galaxy clusters
(Table~\ref{data1:Kopylova_n_en}), they have similar radial velocities.

{\it The A\,1190 supercluster} is a part of the SCl\,38 supercluster.
In the catalog of galaxy superclusters~\cite{Einasto1:Kopylova_n_en} and earlier
catalogs, five galaxy clusters have been found in this region:
A\,1155, A\,1173, A\,1187, A\,1190, A\,1203. We have determined
parameters of these systems and found that the clusters A\,1155 and
A\,1187 correspond to poor galaxy groups with the radial velocity
dispersion smaller than  200~km\,s$^{-1}$, thus, we do not consider
them here. Actually, there are three rich galaxy clusters in this
supercluster,  A\,1190 is the most massive system of all.

{\it The A\,1750/A\,1809 supercluster}  is a part of the  SCl\,61
supercluster. We have determined the peculiar motions only for the
most massive galaxy systems in this supercluster~\cite{Einasto1:Kopylova_n_en}. All
three superclusters have close redshifts and are characterized by the
presence of several peaks in the  radial velocity dispersions of the
clusters belonging to them, consequently, they are forming systems.

Apart from large galaxy superclusters, we also considered the pairs
of the galaxy clusters: A\,2029 and A\,2033,  A\,2244 and A\,2245.

\subsection{Selection of Early-Type Galaxies}

In the present paper, to determine relative distances of galaxy
clusters we used galaxy parameters given in the SDSS Data Release
8~\cite{Aihara:Kopylova_n_en}. Using the data of this issue, which takes
into consideration  the errors in reducing massive galaxy images made
in previous issues, we built the FP of 93\,000 early-type galaxies
with $z<0.2$  in all filters~\cite{Saulder:Kopylova_n_en}. We used the shape of the
built FP having specified the zero-point for our sample along with
this. While selecting early-type galaxies in the groups and clusters
under study, we used the following criteria which slightly differ
from those in paper~\cite{Saulder:Kopylova_n_en}:
%\begin{minipage}{1.0\linewidth}
\begin{list}{}{
\setlength\leftmargin{2mm} \setlength\topsep{2mm}
\setlength\parsep{0mm} \setlength\itemsep{2mm} }
 \item 1) the central stellar velocity dispersion---$100<\sigma<420$~km\,s$^{-1}$;
 \item 2) the parameter characterizing the contribution of the de Vaucouleurs profile into the surface-brightness profile---$\rm fracDeV_r\geq 0.8$;
 \item 3) the concentration index equal to the ratio of the radii containing 90\% and 50\% of the Petrosian fluxes---$r_{90}/r_{50}\geq2.6$;
 \item 4) the restriction by color---$\Delta (u-r)>-0.2$ in order to exclude late-type galaxies, where\linebreak \mbox{$\Delta (u-r)$ = $(u-r)+0.108\,M_r-0.3$}~\cite{Kopylova4:Kopylova_n_en};
 \item 5) the relation of axes of the galaxies---\mbox{$\rm deVAB\geq 0.3$};
 \item 6) the $S/N$ ratio in the galaxy spectra--- $\rm snMedian>10$;
 \item 7) the limiting value for our sample which corresponds to the spectral limit of the SDSS equal to the value not corrected for absoption, \mbox{$r_{\rm Pet}=17\fm77$}~\cite{Strauss:Kopylova_n_en}.
\end{list}
%\end{minipage}

The quantity of the galaxies used plays a key role in determination
of relative distances of galaxy clusters, as the standard error of
the average distance equals the standard deviation divided by
$\sqrt{(N)}$. However, it is important that the galaxies are selected
homogeneously. Our  criteria are close to those applied
in~\cite{Pearson:Kopylova_n_en}. Our main criteria are~(1) and~(2). Because of the
galaxies with  \mbox {$\sigma<100$}~km\,s$^{-1}$, the scatter in the
distances to determine increases~\cite{Jorgensen:Kopylova_n_en}. Accuracy of
measurements of such $\sigma$ is lower than accuracy of measurements
of high velocities. Criterion~(4) makes it possible to exclude
late-type galaxies left after applying criterion~(2). The other
criteria provide a stricter selection of early-type galaxies.
Moreover, with one and the same radial velocity dispersion $\sigma$,
galaxies can have slight population differences (metallicity,
age)~\cite{Jorgensen:Kopylova_n_en}. We excluded such galaxies by their residual
deviations from the FP (Fig.~3, item 2.4).

We used galaxy parameters that were obtained by fitting the de
Vaucouleurs profile to the observed galaxy profile. All corrections:
aperture corrections~
($\sigma_0=\sigma_{\rm sdss}(r_{\rm fiber}/ (r_{\rm cor}/8))^{0.04}$,
where  \mbox{$r_{\rm
cor}=r_{\rm dev}\,\sqrt{(b/a)}$} is the model radius of a galaxy from
the SDSS with regard to its ellipticity), corrections for absorption
in a Galaxy (from the SDSS data), $K$
correction~\cite{Chilingarian:Kopylova_n_en}---were made in
accordance with paper~\cite{Saulder:Kopylova_n_en}. Radial velocities
of galaxy clusters were normalized to the system of the CMB:
corrections were taken from the NED data base.
In the correction for cosmological dimming of the average surface
brightness
$$
\begin{array}{rcl}
\mu_e&=&m_{\rm dev}+ 2.5\,\log\,(2\pi\,r_{\rm dev}^2)\\
&-&K(z)-5\,\log\,(1+z_{\rm ph})- 5\,\log\,(1+z_{\rm sp})
\end{array}
$$
it has been taken into account that $z_{\rm sp}$ is the measured
redshift which also includes the peculiar velocity of an object and
$z_{\rm ph}$ is the redshift corresponding to the actual cosmological
distance~\cite{Mohr:Kopylova_n_en}. When estimating the surface brightness, only
the first part of the correction was taken into account, the second
part of it was taken into account in the zero point in determination
of the Hubble relation (Section~2.4)

\subsection{The Fundamental Plane}
\begin{figure}[]
\setcaptionmargin{5mm} \onelinecaptionsfalse
\includegraphics[scale=0.5,angle=0]{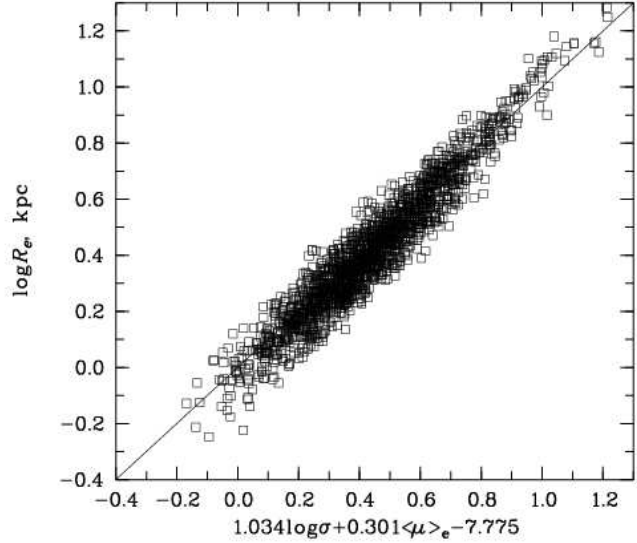}
\captionstyle{normal} \caption{Projected FP of early-type galaxies
along the long axis, effective radius, for galaxy clusters. The line
corresponds to the common zero point of the sample
\mbox{$\gamma=-7.775$}. } \label{FP:Kopylova_n_en}
\end{figure}
Figure~\ref{FP:Kopylova_n_en} shows the FP derived in the  $r$ band (SDSS) for 1732
early-type galaxies selected according to the above criteria. The
line corresponds to the forward regression~\cite{Saulder:Kopylova_n_en} relative to
$\log R_e$   and is written as:
\begin{equation}
\log R_e\,[{\rm kpc}]=1.034\,\log \sigma+0.301\,\langle\mu
\rangle_e+\gamma,
\end{equation}
where $R_e$ is the effective radius of a galaxy in kpc, $\langle\mu_e
\rangle$ is the average surface brightness within the limits of this
radius, and  $\sigma$ is the central dispersion of stellar radial
velocities. We determined the more accurate zero point $\gamma$ for
our sample and obtained  $\gamma=-7.775$. The root-mean-square
deviation of the FP zero point is equal to $0.068\pm0.003$ which is
equivalent to the 16\% error of the distance determination for a
single galaxy. The formal error  of the cluster distance
determination depends on the number of the used galaxies and varies
from  2\% to 12\%. Measurement errors of the FP galaxy parameters
give the contribution to the root-mean-square deviation on the 0.045
FP (the SDSS data~\cite{Batiste:Kopylova_n_en}). Age and metallicity of galaxies
can contribute additional scattering; they can be determined from the
Mg$_2$ line, although, according to~\cite{Hudson:Kopylova_n_en}, while comparing
residual deviations from the FP and from the relation
Mg$_2$--$\sigma$, we have not found any strong correlation between
them. Inconsiderable influence of the  $E$ and $SO$ galaxies
separately on the coefficients and the FP zero point was
found~\cite{Hudson:Kopylova_n_en}. Moreover, the galaxy scatter on the FP is
increased by subsystems belonging to the clusters under
study~\cite{Gibbons:Kopylova_n_en}. The homogeneous Malmquist bias significantly
contributes into the galaxy scatter on the FP; we introduced the
correction for it according to~\cite{Kaiser:Kopylova_n_en}. The zero point varies
with distance of galaxies in case $\log R_e$ varies in arcseconds, as
the other parameters ($\langle\mu_e \rangle$ and $\sigma$) do not
vary with distance. According to~\cite{Kaiser:Kopylova_n_en}, when using the
forward FP for estimation of distances, corrections for the
homogeneous Malmquist bias should be taken into consideration
(equation (A3) in this paper). We took this bias into account when
determining distances of galaxy clusters.

\begin{figure*}[]
\setcaptionmargin{5mm} \onelinecaptionsfalse
\includegraphics[scale=0.5,angle=0]{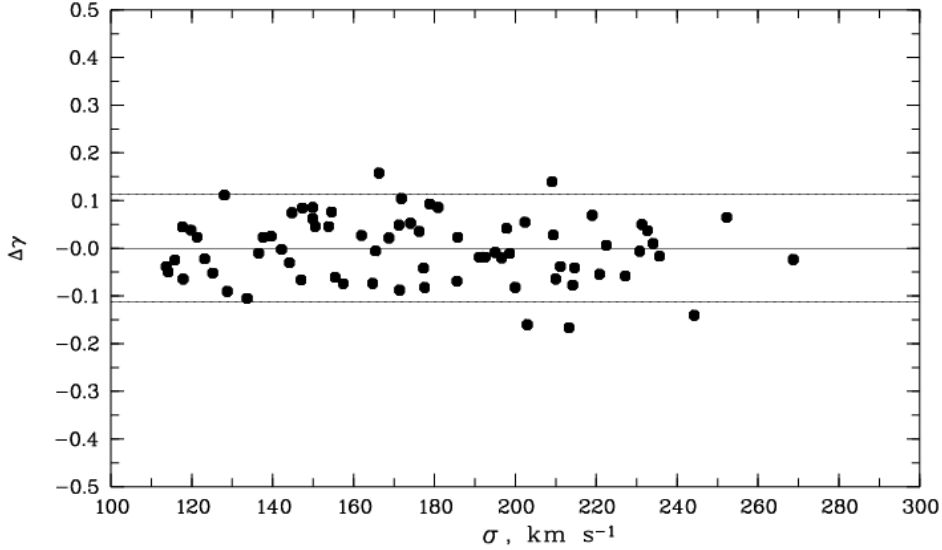}
\captionstyle{normal} \caption{Residual deviations of the FP $\Delta
\gamma$ versus the central stellar dispersion
$\sigma$ for the A\,2065 cluster. Horizontal lines correspond to the
deviations from the average $\gamma$ at the level of $\pm2\sigma$.}
\label{A2065:Kopylova_n_en}
\end{figure*}

Residual deviations from the FP
$$
\Delta \gamma = \log R_e\,[{\rm arcsec}]-
1.034\,\log\sigma-0.301\,\langle\mu_e \rangle-\gamma
$$
do not depend on the central stellar dispersion in galaxies. We used
this fact to improve the sample of the galaxies selected earlier in
each cluster (Fig.~\ref{A2065:Kopylova_n_en}). Empirically, we found that almost
all deviations of the~$\gamma$ zero points of galaxies from the
average zero point of the cluster do not exceed $2\sigma$. The figure
shows the A\,2065 cluster as an example, in which galaxies beyond
$2\sigma$ (horizontal lines in the figure) were not considered while
calculating the average zero point of the cluster.

\section{PECULIAR VELOCITIES OF GALAXY CLUSTERS}
\begin{figure*}[]
\setcaptionmargin{5mm} \onelinecaptionsfalse
\includegraphics[scale=0.5,angle=0]{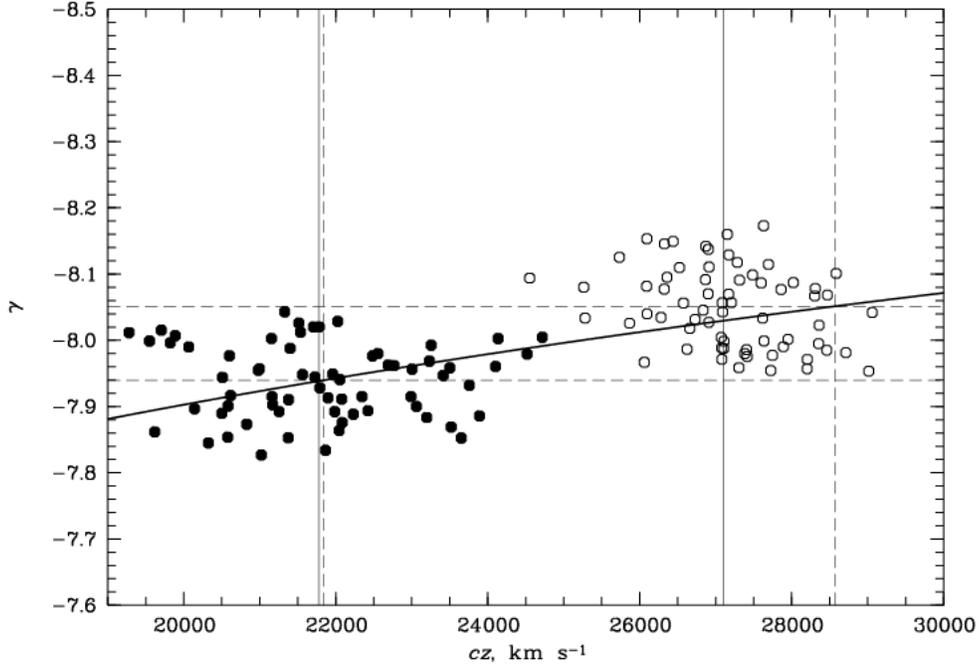}
\captionstyle{normal} \caption{Dependence of the individual
distances, zero points of the FP $\gamma$ on the radial velocities of
the galaxies located within the radius  $R_{200}$ in the clusters
A\,2065 (solid circles) and A\,2142 (open circles). A bold line
corresponds to the Hubble relation between radial velocity and
distance. Solid vertical lines show average radial velocities of
clusters  $cz_{\rm CMB}$ that give corresponding distances at the
inter-crossing with the Hubble curve; dashed lines show average
distances of the clusters found with the FP and radial velocities
corresponding to them $cz_{\rm FP}$.} \label{gamma:Kopylova_n_en}
\end{figure*}

Figure~\ref{gamma:Kopylova_n_en} gives the observed individual
distances (the $\gamma$ zero points calculated with $\log R_e$ in
arcseconds) of the galaxies in the  A\,2065  cluster (solid circles)
and in the more remote cluster A\,2142 (open circles) depending on
their radial velocities relative the CMB. A bold
line shows the expected Hubble dependence between radial velocity and
distance calculated for a model we accepted and the zero point equal
to $-7.775$. In this  case, the comoving radial
distance\linebreak \mbox{$D=(cz/H_o)\,(1-0.225z)/(1+z)$} is
calculated with $q_o=-0.055$, and the correction for stellar
evolution in the galaxy \mbox{$Q=1.07z$} is introduced in the zero
point of the FP~\cite{Saulder:Kopylova_n_en} and a part of geometric
dimming of light of the galaxies is taken into account
$5\,\log\,(1+z_{\rm ph}$). Solid vertical lines near each cluster
show their radial velocities  relative to the CMB
determined for the region with the radius $R_{200}$. Dashed
(horizontal) lines show average distances of the galaxy clusters
determined from the FP and radial velocities corresponding to them
(vertical lines). The difference of radial velocities is
characterized as a peculiar velocity of a group or a galaxy cluster
along the line of sight. In other words,  $V_{\rm pec}=c\,(z_{\rm
CMB}-z_{\rm FP})/(1+z_{\rm FP})$, where $c$ is the light velocity,
$c\,z_{\rm CMB}$ is the radial velocity of a cluster relative to the
CMB, and  $c\, z_{\rm FP}$ is the radial velocity of
a cluster corresponding to a distance determined with the FP.
Table~\ref{data2:Kopylova_n_en} presents the results obtained for the
galaxy systems under study. The first column contains the name of the
galaxy system; the second---the average distance of the system (the
observed zero point $\gamma$  with error); the third---the number of
galaxies used; the fourth shows the redshift  $z_{\rm FP}$
corresponding to $\gamma$; the fifth---the distance  $D$  in Mpc
corresponding to $z_{\rm FP}$; the sixth column gives the peculiar
velocity with error; the seventh---the relation of the peculiar
velocity to its error. Table~\ref{data3:Kopylova_n_en} presents
average peculiar velocities of the galaxy cluster systems and their
rms deviations along the line of sight with squared error check.

%2
\onecolumngrid
\setcaptionwidth{\linewidth}%
\setcaptionmargin{0mm} %
\onelinecaptionstrue \captionstyle{normal}
\medskip
\begin{longtable*}{l|c|c|c|c|r@{$\,\pm\,$}l|c}
\caption{Peculiar velocities of galaxy groups and clusters}\label{data2:Kopylova_n_en}\\
\hline
\multicolumn{1}{c|}{Cluster} & $\gamma$ &  $N_{\rm FP}$ & $z_{\rm FP}$& $D$, Mpc    & \multicolumn{2}{c|}{$V_{\rm pec}$,  km\,s$^{-1}$} & $V_{\rm pec}/\Delta V_p$ \\
\hline
\multicolumn{1}{c|}{(1)}     &  (2)    &  (3)     & (4)& (5)   & \multicolumn{2}{c|}{(6)}        &  (7) \\
\hline
\endfirsthead
\caption{Continued}\\
\hline
\multicolumn{1}{c|}{Cluster} & $\gamma$ &  $N_{\rm FP}$ & $z_{\rm FP}$& $D$, Mpc    & \multicolumn{2}{c|}{$V_{\rm pec}$,  km\,s$^{-1}$} & $V_{\rm pec}/\Delta V_p$ \\
\hline
\multicolumn{1}{c|}{(1)}     &  (2)    &  (3)     & (4)& (5)   & \multicolumn{2}{c|}{(6)}        &  (7) \\
\hline
\endhead
\hline
\endfoot
\endlastfoot
SCl CrB             &                  &    &        &       & \multicolumn{2}{c|}{} &      \\
A\,2019             & $-7.958\pm0.016$ &  9 & 0.0760 & 319.9 & $1591$&$810$   & 1.96 \\
A\,2056             & $-7.898\pm0.032$ &  3 & 0.0660 & 278.5 & $2645$&$1530$  & 1.73 \\
A\,2061             & $-7.945\pm0.009$ & 40 & 0.0739 & 311.1 & $1197$&$470$   & 2.55 \\
A\,2065             & $-7.940\pm0.007$ & 66 & 0.0728 & 306.9 & $-58$&$340$    & 0.17 \\
A\,2067             & $-7.956\pm0.008$ &  7 & 0.0758 & 319.3 & $-700$&$420$   & 1.67 \\
A\,2079             & $-7.891\pm0.012$ & 24 & 0.0649 & 273.9 & $347$&$520$    & 0.67 \\
A\,2089             & $-7.929\pm0.014$ & 17 & 0.0710 & 299.3 & $814$&$650$    & 1.25 \\
A\,2092             & $-7.899\pm0.015$ & 15 & 0.0662 & 279.3 & $205$&$680$    & 0.30 \\
A\,2124             & $-7.915\pm0.012$ & 32 & 0.0687 & 289.5 & $-747$&$550$   & 1.36 \\
CL\,1529+29         & $-7.980\pm0.008$ & 18 & 0.0801 & 337.5 & $1203$&$900$   & 1.34 \\
\hline
A\,2142             & $-8.052\pm0.007$ & 67 & 0.0953 & 399.5 & $-1343$&$510$  & 2.63 \\
\hline
SCl Bootes          &                  &    &        &       & \multicolumn{2}{c|}{} &    \\
A\,1775A            & $-7.895\pm0.037$ &  6 & 0.0655 & 276.6 & $241$&$1720$   & 0.14\\
A\,1781             & $-7.910\pm0.019$ & 12 & 0.0678 & 281.7 & $-1349$&$910$  & 1.48\\
A\,1795             & $-7.930\pm0.003$ & 29 & 0.0712 & 299.9 & $-2223$&$540$  & 4.12\\
A\,1825             & $-7.862\pm0.018$ & 11 & 0.0605 & 255.7 & $1006$&$780$   & 1.29\\
A\,1828             & $-7.881\pm0.014$ &  7 & 0.0634 & 267.5 & $12$&$640$     & 0.02\\
A\,1831A            & $-7.858\pm0.019$ & 10 & 0.0599 & 253.2 & $1063$&$820$   & 1.30\\
A\,1775B            & $-7.966\pm0.009$ & 27 & 0.0776 & 326.4 & $-446$&$500$   & 0.89\\
A\,1800             & $-7.977\pm0.010$ & 30 & 0.0797 & 335.2 & $-999$&$570$   & 1.75\\
A\,1831B            & $-7.941\pm0.009$ & 34 & 0.0731 & 307.8 & $881$&$500$    & 1.76\\
A\,1898             & $-7.944\pm0.024$ &  7 & 0.0735 & 309.7 & $1586$&$1290$  & 1.23\\
CL\,1350+29         & $-7.994\pm0.021$ & 13 & 0.0830 & 348.9 & $-1615$&$1220$ & 1.32\\
\hline
SCl Z\,5029/A\,1424 &                  &    &        &       & \multicolumn{2}{c|}{} &   \\
A\,1424             & $-7.945\pm0.011$ & 30 & 0.0738 & 310.6 & $931$&$580$    & 1.60\\
A\,1516             & $-7.948\pm0.012$ & 26 & 0.0744 & 313.2 & $820$&$650$    & 1.26\\
Z\,4905             & $-8.001\pm0.014$ & 25 & 0.0845 & 354.9 & $-1794$&$1040$ & 1.72\\
Z\,5029             & $-7.980\pm0.009$ & 39 & 0.0802 & 337.2 & $-461$&$520$   & 0.89\\
\hline
SCl A\,1190         &                  &    &        &       & \multicolumn{2}{c|}{} &    \\
A\,1173             & $-7.984\pm0.016$ & 20 & 0.0810 & 340.7 & $-1112$&$960$  & 1.16\\
A\,1190             & $-7.979\pm0.011$ & 38 & 0.0801 & 337.0 & $-1098$&$630$  & 1.74\\
A\,1203             & $-7.952\pm0.013$ & 12 & 0.0751 & 316.2 & $289$&$690$    & 0.37\\
\hline
SCl A\,1750/A\,1809 &                  &    &        &       & \multicolumn{2}{c|}{} &     \\
A\,1750             & $-8.006\pm0.008$ & 45 & 0.0854 & 358.8 & $418$&$490$    & 0.85\\
A\,1773             & $-7.940\pm0.013$ & 30 & 0.0730 & 307.4 & $1515$&$680$   & 2.23\\
A\,1780             & $-7.946\pm0.025$ & 14 & 0.0740 & 311.5 & $1352$&$1320$  & 1.02\\
A\,1809             & $-7.976\pm0.010$ & 37 & 0.0794 & 334.0 & $223$&$550$    & 0.40\\
\hline
A\,0602             & $-7.870\pm0.017$ & 14 & 0.0617 & 260.5 & $-142$&$740$   & 0.19\\
A\,0671             & $-7.783\pm0.016$ & 30 & 0.0497 & 210.5 & $238$&$590$    & 0.40\\
A\,1024             & $-7.964\pm0.014$ & 14 & 0.0772 & 325.0 & $-762$&$800$   & 0.95\\
A\,1035A            & $-7.913\pm0.010$ & 37 & 0.0683 & 288.1 & $143$&$480$    & 0.30\\
A\,1035B            & $-7.954\pm0.018$ & 17 & 0.0754 & 317.4 & $1014$&$910$   & 1.11\\
A\,1066             & $-7.939\pm0.012$ & 25 & 0.0728 & 306.8 & $-747$&$600$   & 1.24\\
A\,1205             & $-7.992\pm0.013$ & 27 & 0.0833 & 349.9 & $-1645$&$770$  & 2.14\\
A\,1238             & $-7.946\pm0.010$ & 16 & 0.0740 & 311.4 & $386$&$560$    & 0.68\\
A\,1371             & $-7.913\pm0.014$ & 20 & 0.0684 & 288.3 & $465$&$710$    & 0.65\\
A\,1589             & $-7.938\pm0.012$ & 34 & 0.0726 & 305.7 & $-28$&$610$    & 0.04\\
A\,1668             & $-7.898\pm0.011$ & 25 & 0.0660 & 278.3 & $-363$&$520$   & 0.70\\
A\,1691             & $-7.925\pm0.008$ & 40 & 0.0703 & 296.7 & $832$&$380$    & 2.19\\
A\,1767             & $-7.940\pm0.009$ & 30 & 0.0730 & 307.5 & $-641$&$480$   & 1.34\\
A\,1904             & $-7.931\pm0.006$ & 40 & 0.0714 & 301.0 & $200$&$300$    & 0.67\\
A\,1991             & $-7.832\pm0.011$ & 28 & 0.0564 & 238.5 & $756$&$450$    & 1.68\\
A\,2029             & $-7.980\pm0.009$ & 61 & 0.0803 & 337.8 & $-518$&$410$   & 0.75\\
A\,2033             & $-7.989\pm0.015$ & 26 & 0.0820 & 344.6 & $-372$&$900$   & 0.41\\
A\,2064             & $-7.964\pm0.022$ & 13 & 0.0773 & 325.2 & $-921$&$1230$  & 0.75\\
A\,2244             & $-8.063\pm0.007$ & 33 & 0.0981 & 410.9 & $229$&$510$    & 0.45\\
A\,2245             & $-7.979\pm0.008$ & 37 & 0.0800 & 336.6 & $2185$&$470$   & 4.65\\
RXC\,J1022          & $-7.825\pm0.014$ & 19 & 0.0554 & 234.5 & $-127$&$580$   & 0.22\\
RXC\,J1351          & $-7.870\pm0.013$ & 24 & 0.0617 & 260.7 & $493$&$570$    & 0.86\\
SHK\,352            & $-7.796\pm0.015$ & 23 & 0.0518 & 219.2 & $-409$&$550$   & 0.74\\
Z\,6718             & $-7.938\pm0.017$ & 12 & 0.0726 & 305.8 & $-225$&$910$   & 0.25\\
\hline
%\end{tabular}
%\end{table*}
\end{longtable*}
%\twocolumngrid

\begin{figure}[]
\setcaptionmargin{5mm} \onelinecaptionsfalse \hspace*{-1cm}
\includegraphics[scale=0.5,angle=0]{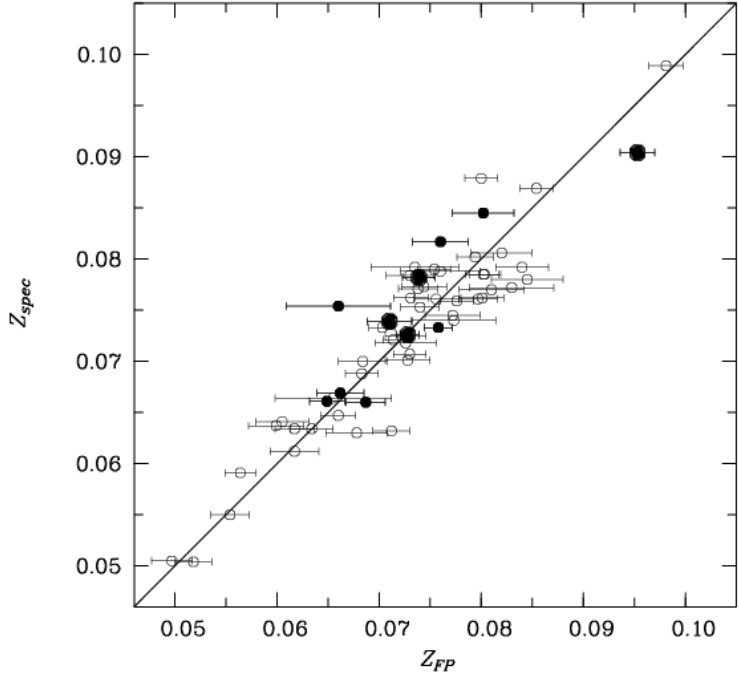}
\captionstyle{normal} \caption{The Hubble diagram (velocity---$z_{\rm
spec}$, distance---$z_{\rm FP}$) for the region of the Corona
Borealis galaxy supercluster: the supercluster members A\,2061,
A\,2065, and A\,2089 are shown with big solid circles and
others---with small solid circles. Open circles denote the whole
sample under study. } \label{XCrB:Kopylova_n_en}
\end{figure}

Figure~\ref{XCrB:Kopylova_n_en} shows the Hubble diagram for galaxy systems
in the region of
the Corona Borealis supercluster and the A\,2142 cluster
against the background of the whole sample. Earlier, in
Fig.~\ref{clus1:Kopylova_n_en}, we show that the  A\,2067 system is a small group
with the radial velocity dispersion smaller than  300~km\,s$^{-1}$.
The A\,2056 system is also a poor group with a small number of
members (Table~\ref{data1:Kopylova_n_en}). If we do not consider them, then the
average peculiar velocity of A\,2061+A\,2065+A\,2089, the
gravitationally bound core of the supercluster~\cite{Pearson:Kopylova_n_en}, is
positive and equal to  $651\pm290$~km\,s$^{-1}$. The galaxy clusters
move away as for the CMB with a velocity
2.2~times exceeding the measurement error. The rms deviation of
their peculiar velocities is  \mbox
{$668\pm150$}~km\,s$^{-1}$~(Table~3).

However, as the richest cluster in CrB A\,2065 does not have a
peculiar velocity  (Table~\ref{data2:Kopylova_n_en}), then one can suppose that the
most remote clusters in  CrB with great positive peculiar velocities,
A\,2019, A\,2061, and CL\,1529+29, move toward the massive galaxy
supercluster having A\,2142  along the filament connecting
them~\cite{Pillastrini:Kopylova_n_en}. It was found using the  SDSS DR8 and DR10
data that this supercluster with the center in A\,2142 is a
supercluster with the collapsing core~\cite{Einasto1:Kopylova_n_en}.

%3
\begin{table*}[]
\setcaptionmargin{0mm} \onelinecaptionstrue \captionstyle{normal}
\caption{Peculiar velocities and rms deviations of peculiar
velocities in galaxy clusters} \label{data3:Kopylova_n_en}
\medskip
\begin{tabular}{l|c|r|r@{$\,\pm\,$}l|c}
\hline
\multicolumn{1}{c|}{Sample} &   $N$  & \multicolumn{1}{c|}{$\langle V_{\rm pec} \rangle$}  & \multicolumn{2}{c|}{$\langle V_{\rm pec}^2 \rangle^{1/2}$} & Ref.\\
\hline
SCl CrB                   &  9 & $428\pm210$   & $652$&$50$   & this work\\
A\,2061, A\,2065, A\,2089 &  3 & $651\pm290$   & $668$&$150$  & this work\\
SCl Boo                   & 11 & $-168\pm280$  & $757$&$70$   & this work\\
                          &  6 & $-208\pm400$  & $728$&$150$  & this work\\
                          &  5 & $-118\pm400$  & $791$&$120$  & this work\\
SCl Z\,5029/A\,1424       &  4 & $535\pm360$   & $1366$&$170$ & this work\\
SCl A\,1190               &  3 & $-640\pm450$  & $496$&$130$  & this work\\
SCl A\,1750/A\,1809       &  4 & $877\pm420$   & $630$&$170$  & this work\\
SCl Her                       & 13 & $-72\pm240$   & $736$&$50$   &\cite{Kopylova3:Kopylova_n_en} \\
SCl Leo                       & 13 & $168\pm320$   & $625$&$70$   &\cite{Kopylova3:Kopylova_n_en} \\
A\,2244+A\,2245           &  2 &               & $1473$&$540$ & this work\\
A\,2029+A\,2033           &  2 & $-445\pm700$  & $0$&$170$    & this work\\
field                     & 20 & $-59\pm140$   & $0$&$20$     & this work\\
total                     & 57 & $83\pm130$    & $646$&$10$   & this work\\
\hline
\end{tabular}
\end{table*}

A\,2142 is a massive cluster (according our estimates, its mass
$M_{200} = 1.5\times 10^{15}$~$M_{\odot}$) at $z=0.09$ with the
highest X-ray luminosity~\cite{Ebeling:Kopylova_n_en} in the whole sample.
Figure~\ref{gamma:Kopylova_n_en} shows it with open circles and it can be noted
that the basic amount of galaxies are situated further ($\gamma$ is
smaller) than they should according to the Hubble dependence (a bold
line). The A\,2142 cluster has a negative peculiar velocity 2.6~times
exceeding the error, i.e, the cluster is possibly moving toward the
supercluster to CrB. Thus, the gravitational interaction of large
galaxy cluster systems of the CrB supercluster and the supercluster
with A\,2142~\cite{Pillastrini:Kopylova_n_en} is confirmed.

The A\,2079, A\,2092, and A\,2124 clusters show small peculiar
velocities; most probably  they are not significantly influenced by
the central part of the cluster. On the whole,  within CrB we observe
considerable peculiar motions along the line of sight with the rms
deviation \mbox {$652\pm50$}~km\,s$^{-1}$ which considerably exceeds
the motions of galaxy clusters not belonging to big structures,
$0\pm20$~km\,s$^{-1}$. In the earlier paper~\cite{Kopylova1:Kopylova_n_en}, using
Kormendy relation (two-parameter), we found significant negative
peculiar velocities in the A\,2089 and A\,2092 galaxy clusters and,
as a result, inferred on the gravitational collapse of the
supercluster's core. In the present work performed with the FP (a
three-parameter plane) and a large number of galaxies, we do not find
such velocities in these galaxy systems. In other galaxy clusters,
the signs of peculiar velocities determined with two methods
coincide, except for the poor A\,2067 group. If we compare the
peculiar velocities found and the results of paper~\cite{Pearson:Kopylova_n_en}
from the  SDSS (DR7) data, then we can note that the signs of
peculiar velocities coincide in all the galaxy clusters except for
the systems A\,2067 and A\,2092.

\begin{figure}[]
\setcaptionmargin{5mm} \onelinecaptionsfalse
\includegraphics[scale=0.5,angle=0]{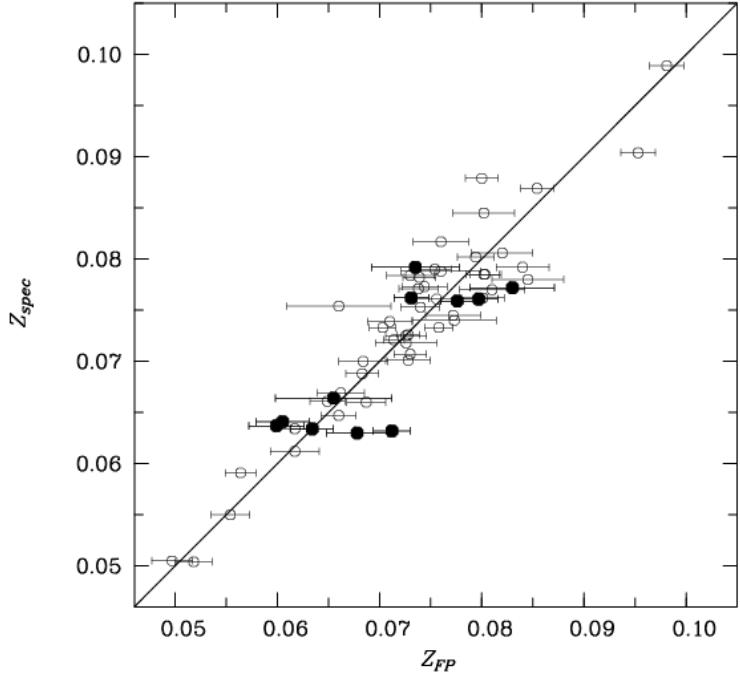}
\captionstyle{normal} \caption{The same as in Fig.~\ref{XCrB:Kopylova_n_en} but for
the region of the Bootes supercluster.} \label{XBoo:Kopylova_n_en}
\end{figure}

Figure~\ref{XBoo:Kopylova_n_en} shows the Hubble diagram for the region of the
Bootes supercluster. The supercluster consists of two galaxy
superclusters projected on each other and having the difference of
peculiar velocities of about  4000~km\,s$^{-1}$. It can be noted that
peculiar motions in both systems are significant and their rms
deviations are greater than those for the CrB supercluster. The
peculiar velocity of the majority of the clusters inconsiderably
exceeds the accuracy of measurements ($V_{\rm pec}/\Delta V_p <1.8$).
The peculiar velocity of the richest cluster in the nearby A\,1795
subsystem appeared unexpectedly high; the cluster has a mass of
$7.8\times 10^{14}$~$M_{\odot}$ within  $R_{200}$. It is equal to
about 2223~km\,s$^{-1}$ and four times exceeds the measurement error.
A\,1795 is a powerful X-ray source~\cite{Ebeling:Kopylova_n_en}, contains an active
cD galaxy with a peculiar velocity of 245~km\,s$^{-1}$ (relative to
the average velocity of the cluster), and has two more peaks in the
radial velocity dispersion apart from the central one with the
cD~galaxy.

The greatest rms deviations $\langle V_{\rm pec}^2 \rangle^{1/2}$ of
one-dimensional peculiar velocities are observed in the cluster
system A\,1750/A\,1809 and in the pair of rich galaxy clusters
A\,2244 and A\,2245. The A\,2244 cluster is the major in this pair,
as it has the X-ray  luminosity an order larger than that for
A\,2245~\cite{Ebeling:Kopylova_n_en}. With a radial velocity difference between
these clusters of 3310~km\,s$^{-1}$ and a distance between them of
about 45~Mpc, the  A\,2245 cluster, being situated closer to us than
the  A\,2244 cluster, is moving toward it  with a great positive
peculiar velocity. In A\,2245, there is a dip  in the radial velocity
dispersion, i.e., it consists of two subsystems with a difference in
radial velocities of about  1714~km\,s$^{-1}$. We found that both
subsystems have positive peculiar velocities. In the pair of the
A\,2029 and A\,2033 galaxy clusters, $\langle V_{\rm pec}^2
\rangle^{1/2}$ is equal to zero. The major system is A\,2029 with
very high X-ray luminosity~\cite{Ebeling:Kopylova_n_en} and  A\,2033 is a smaller
system gravitationally bound with it. Both clusters are situated at
the same distance $\gamma$ and their small peculiar velocities have
the same sign; they are moving away from us. The whole sample of
galaxy clusters has an inconsiderable peculiar velocity relative to
the CMB \mbox {$83\pm130$}~km\,s$^{-1}$. The rms
deviation $\langle V_{\rm pec}^2 \rangle^{1/2}$ of peculiar
velocities of 20 galaxy clusters beyond the massive structures with
\mbox{$\sigma>500$~km\,s$^{-1}$} is equal to
\mbox{$0\pm20$~km\,s$^{-1}$}; if we take the same clusters in the
superclusters, then we will obtain the rms deviation  \mbox
{$876\pm30$}~km\,s$^{-1}$.

\section{CONCLUSIONS}

Model calculations show that high peculiar velocities, $V_{\rm
pec}>10^3$~km\,s$^{-1}$, appear in dense galaxy
superclusters~\cite{Bahcall:Kopylova_n_en}. In the present paper, we
analyze the dynamics of galaxy clusters in the rich superclusters
Corona Borealis and Bootes and in poorer superclusters
Z\,5029/A\,1424, A\,1190, A\,1750/A\,1809, in pairs and galaxy
clusters which do not belong to any large structures and do not have
neighbors comparable in size within the limits of about 4~Mpc  in the
sky plane in the radial velocity range of
$\pm$~3000~km\,s$^{-1}$. The work has been done using the  SDSS (DR8)
and NED databases. The CrB and Bootes galaxy superclusters  are
extended structures projected on the celestial sphere that occupy
approximately \mbox{$50\times 45$}~Mpc and $50\times 53$~Mpc. Within
these limits, ten and eleven galaxy systems respectively are located
in the projection; they extend to the third coordinate at 76 and
68~Mpc. In order to find peculiar motions of galaxy clusters (57
systems), we determined their relative distances with the FP of the
early-type galaxies and compared them with the distances found from
the Hubble's law.

The peculiar velocities that we obtained vary in module from 12 to
2200~km\,s$^{-1}$. Five galaxy clusters have peculiar velocities more
than two times exceeding measurement errors (A\,2061, A\,2142,
A\,1773, A\,1205, A1691), two clusters, A\,1795 and A\,2245,---four
times. The rms deviation of peculiar velocities is maximum in the
A\,1750/\,1809 supercluster and in the pair of the clusters A\,2244
and A\,2245. In the CrB supercluster, the rms deviation of peculiar
velocities is also large and equals  \mbox {$652\pm50$}~km\,s$^{-1}$,
and in the Boo supercluster components---\mbox
{$728\pm150$}~km\,s$^{-1}$ and $791\pm120$~km\,s$^{-1}$. In the
Hercules and Leo galaxy superclusters, we earlier obtained similar
results using the SDSS (DR7) data~\cite{Kopylova3:Kopylova_n_en}. Table~\ref{data3:Kopylova_n_en}
presents these results. In the CrB supercluster, the massive cluster
A\,2065 has no peculiar velocity. To explain the observed peculiar
velocities of other members of galaxy systems, there are two
variants: the collapsing central region of  CrB consisting of
A\,2056, A\,2061 A\,2065, A\,2067, A\,2089 and, possibly,
A\,2092~\cite{Pearson:Kopylova_n_en} and the motion of the remotest galaxy clusters
A\,2019, A\,2061, and CL\,1529+29 toward the galaxy supercluster with
A\,2142. The A\,2142 cluster that is 50~Mpc distant from the  A\,2061
cluster, in its turn, moves toward the CrB supercluster which
confirms the presence of a gravitational interaction of these galaxy
cluster~systems~\cite{Pillastrini:Kopylova_n_en}.

The main conclusion of our study is that the rms deviation of the
peculiar velocities of galaxy clusters in  superclusters is larger
than 500~km\,s$^{-1}$  and is much higher than the rms deviation of
the peculiar velocities of galaxy systems that do not belong to
massive structures. The mean peculiar velocity of the whole sample of
the galaxy clusters relative to the CMB
is $83\pm130$~km\,s$^{-1}$.

\begin{acknowledgments}
%In the present paper, we used the NASA/IPAC Extragalactic Database
%(NED, \linebreak \url{http://nedwww.ipac.caltech.edu}) and the Sloan
%Digital Sky Survey (SDSS, \url{http://www.sdss.org}).

This research has made use of the NASA/IPAC Extragalactic Database
(NED, \url{http://nedwww.}\linebreak\url{.ipac.caltech.edu}),
which is operated by the Jet Propulsion Laboratory, California Institute of
Technology, under contract with the National Aeronautics and Space
Administration, Sloan Digital Sky Survey (SDSS, \url{http://www.sdss.org}),
which is supported by Alfred P. Sloan Foundation, the participant institutes
of the SDSS collaboration, National Science Foundation, and the United
States Department of Energy.
\end{acknowledgments}

\begin{center}
\refname
\end{center}

\end{document}